# Single-cell approaches to cell competition: high-throughput imaging, machine learning and simulations

Daniel Gradeci[†1,2], Anna Bove[†2,3], Guillaume Charras[*2,3,4],
Alan R. Lowe[*2,4,5], and Shiladitya Banerjee[*1,4]

[1]Department of Physics and Astronomy, University College London, Gower Street, London, WC1E 6BT, UK
[2]London Centre for Nanotechnology, University College London, 17-19 Gordon Street, London, WC1H 0AH, UK
[3]Department of Cell and Developmental Biology, University College London, Gower Street, London, WC1E 6BT, UK
[4]Institute for the Physics of Living Systems, University College London, Gower Street, London, WC1E 6BT, UK
[5]Institute for Structural and Molecular Biology, University College London, Gower Street, London, WC1E 6BT, UK

[†]These authors contributed equally

*correspondence: g.charras@ucl.ac.uk, a.lowe@ucl.ac.uk, shiladitya.banerjee@ucl.ac.uk

**Abstract**
Cell competition is a quality control mechanism in tissues that results in the elimination of less fit cells. Over the past decade, the phenomenon of cell competition has been identified in many physiological and pathological contexts, driven either by biochemical signaling or by mechanical forces within the tissue. In both cases, competition has generally been characterized based on the elimination of loser cells at the population level, but significantly less attention has been focused on determining how single-cell dynamics and interactions regulate population-wide changes. In this review, we describe quantitative strategies and outline the outstanding challenges in understanding the single cell rules governing tissue-scale competition dynamics. We propose quantitative metrics to characterize single cell behaviors in competition and use them to distinguish the types and outcomes of competition. We describe how such metrics can be measured experimentally using a novel combination of high-throughput imaging and machine learning algorithms. We outline the experimental challenges to quantify cell fate dynamics with high-statistical precision, and describe the utility of computational modeling in testing hypotheses not easily accessible in experiments. In particular, cell-based modeling approaches that combine mechanical interaction of cells with decision-making rules for cell fate choices provide a powerful framework to understand and reverse-engineer the diverse rules of cell competition.

# 1. Introduction

## 1.1. What is cell competition?

Competition between cells is a phenomenon that results in the elimination of less fit cells (losers) from a tissue, that is then taken over by cells with greater fitness (winners) [1-3]. For the loser cells, the outcome depends strongly on context: when they are cultured in a homotypic environment, they thrive but, when they are cultured in a mixed population, they are eliminated by the winner cells (**Fig 1A**).

Competition was originally identified in Drosophila, where it was thought to operate as an intrinsic homeostatic control mechanism, maximizing tissue fitness in the growing embryo by destroying suboptimal cells[4, 5]. Initially, most mutations leading to cell competition appeared to confer a loser status to the affected cells (e.g. proteins involved metabolic activity, cell growth, polarity and stress responses). However, mutations that conferred a competitive advantage were soon identified, allowing the affected cells to eliminate normal cells even when vastly outnumbered. Cell competition has now been identified in many species, at different stages of animal development, as well as in physiological and pathological conditions [6].

Over the years, a number of mechanisms of cell competition have been identified, involving either biochemical processes or mechanical interactions (**Fig. 1B**). In all cases, competition has generally been evaluated based on complete elimination of loser cells but significantly less attention has been placed on the dynamics of elimination or determining how cell-scale events lead to population wide changes. In this review, we first provide a short introduction to competition (leaving more detailed descriptions of signaling pathways to other reviews in this issue), then outline the outstanding questions in cell competition at the single-cell scale, and propose quantitative strategies to address these questions.

## 1.2. Competition through biochemical processes

In the broadest sense, biochemical competition can be classified into two major categories: the ligand capture model, which involves a passive competition for a diffusible survival factor, and the comparative fitness model, where cells in contact with one another compare their fitness through signaling[1, 2].

The ligand capture model suggested the existence of a Darwinian-like competition among cells for limiting amounts of pro-survival factors, resulting in the removal of cells with less efficient uptake [7-9]. In this model, loser cells in the vicinity of winners are starved of pro-survival factors but those further away can survive. In the ligand capture model, cell-type specific differences in physico-chemical parameters such as ligand diffusion, ligand-receptor reaction kinetics, and endocytic rate are all likely to play an important role in regulating the outcome of competition.

Subsequent studies, however, have challenged the ligand capture model [10], suggesting an alternative mechanism of competition based on cells sensing their relative fitness. The comparative fitness model postulates that cell competition is governed by short-range cellular interactions. Cells are able to: (i) recognize differences with neighboring cells, (ii) acquire either a winner or loser status in response to the outcome of fitness comparison, (iii) eliminate cells less fit than themselves. This model is based on the definition of fitness as a parameter that cells can measure and compare with one another. Key cellular parameters that have been proposed to control the outcome of comparative fitness competition at the single cell level are the extent and duration of heterotypic cell contact [11], which in turn depend on physical parameters such as intercellular adhesion energy, junctional tension and cortical tension [12].

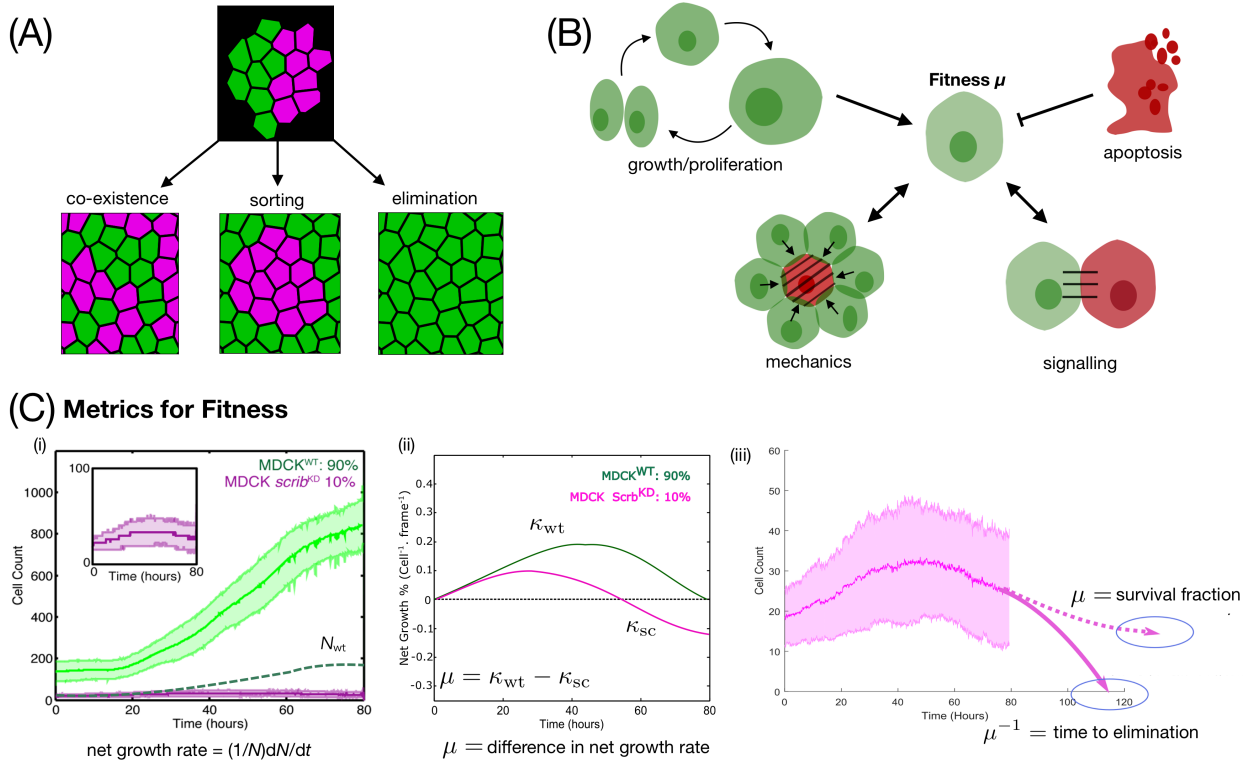

**Figure 1: Cell competition outcomes and strategies.** (A) Cell competition can result in diverse outcomes depending on cell-cell interaction parameters, including stable co-existence, sorting into colonies, or complete elimination of one cell type. (B) Fitness $\mu$ of a cell is increased by factors that increase cell growth/division rate, and is decreased by factors that enhance the rate of apoptosis. Both cell proliferation and apoptosis can be regulated by mechanical forces within the tissue or by short-range biochemical signaling. (C) (i) Fitness for a cell type can be approximated by its net growth rate $\mu = \left(\frac{1}{N}\right) dN/dt$, where $N(t)$ is the cell number at time $t$. (ii) In a competition setting, fitness for a cell type can defined as the difference in its net growth rate from its competitor, i.e. how fast the cell type grows relative to its competitor. (iii) Fitness could also be quantified by the time to elimination of the loser cell type or by the survival fraction.

## 1.3. Mechanical competition

In recent years, experimental observations have suggested the existence of an entirely different mode of cell competition determined by mechanical forces within the tissue. The existence of mechanical competition was first hypothesized by Levayer and colleagues, while studying delamination occurring in the midline of *Drosophila Melanogaster* notum epithelia [13, 14]. In this model, competition acts independently of cell recognition and fitness comparison, rather it results from the combination of faster growth rates in the winner cells and hypersensitivity to crowding in the loser cells [15]. Indeed, experiments *in vitro* and *in vivo* have revealed that cells possess a preferred homeostatic density, for which cell deaths exactly compensate cell births [16]. When crowding is suddenly increased from the homeostatic density, apoptoses and live delaminations decrease cell density [17], while when crowding is decreased suddenly, cell divisions are upregulated [18]. Therefore, if loser cells with low homeostatic density are placed in competition with winner cells with high homeostatic density, apoptoses will be triggered in losers but not in winners once density becomes sufficiently high [19, 20].

The physical or geometrical parameters measured by cells to gain information about their density remain unclear. Recent experiments examining extrusion in homotypic environments suggest that, at high density, live cell extrusion is preceded by increased compression on the extruded cell[21] and that extrusion occurs primarily at locations of defects in the nematic ordering of epithelial cells [22]. In competitive settings, compressive strains also lead to increased probability of extrusion [13]. Both nematic order and cellular strain are affected by physical variables such as cell density, intercellular adhesion, junctional/cortical tension, and cell motility.

**1.4 Quantitative approaches to deciphering the rules of competition**
To date, cell competition has been mostly characterized at the population scale by measuring the elimination of loser cells in movies or still images. Indeed, in its most basic definition, the fitness of a population of cells can be approximated by its net growth rate $\mu=\left(\frac{1}{N}\right)dN/dt$, where $N(t)$ is the number of cells at time $t$ (**Fig. 1C**). A cell type is fitter if it grows more than its competitor (**Fig 1B**). Any factor that increases growth rate will increase fitness, whereas any factor that increases cell death will decrease fitness. Although population scale statistics have been instrumental in describing competition, they do not provide insight into how population shifts arise from the multitude of interactions at the single cell level. Indeed, as with any game, knowing only the result is of little use for determining the rules. Rather, we need to follow the game in detail. In scientific terms, we need to observe each step of the competition, decide on the relevant parameters, characterize each step, and, once we have formulated a reasonable hypothesis, we need to evaluate our understanding by simulations.

In the following sections, we will focus on high-throughput data acquisition, parameterization, characterization, and simulation of cell competition. We will draw on our experience of mechanical competition [20] but similar approaches can be applied to any type of competition.

## 2. Towards a single cell understanding of competition

### 2.1 Single cell-level behaviors in competition
Cell competition is a balance problem that necessitates an understanding of both apoptosis and proliferation. Apoptosis is often quantified in studies of cell competition but faster growing cells may be able to better take advantage of any space freed as a result of cell death [23]. In subconfluent conditions, cell types with higher motility may be able to rapidly disperse to occupy space, whereas, in confluent conditions, mixing allows for more efficient elimination of loser cells [11]. Therefore, at a minimum, we need to characterize cell proliferation, death, and migration.

At a population level, these behaviors can be quantified temporally by computing the number of losers and winners, the fraction of the total population they occupy, their birth/death rate (**Fig 1C**), and their migration velocities or mean square displacement. When competition results in complete elimination of losers, the kinetics of competition can be compared for example by measuring the time to elimination. However, competition may also yield stable coexistence of the two cell types, with losers largely outnumbered but not eliminated (as observed in [11], **Fig 1A**). In this case, a quantification of the survival fraction may allow to compare across conditions (**Fig. 1C**).

To gain insight into competition at the single cell level, we must understand how each of these descriptors of population change is affected by the local environment and, as a consequence, we must also characterize local features of the tissue environment. For example, cell crowding is an important variable in mechanical competition. Therefore, characterizing the local cell density and its temporal evolution is of interest. In comparative fitness competition, contact between winners and losers is a key parameter which should be characterized over time for each cell.

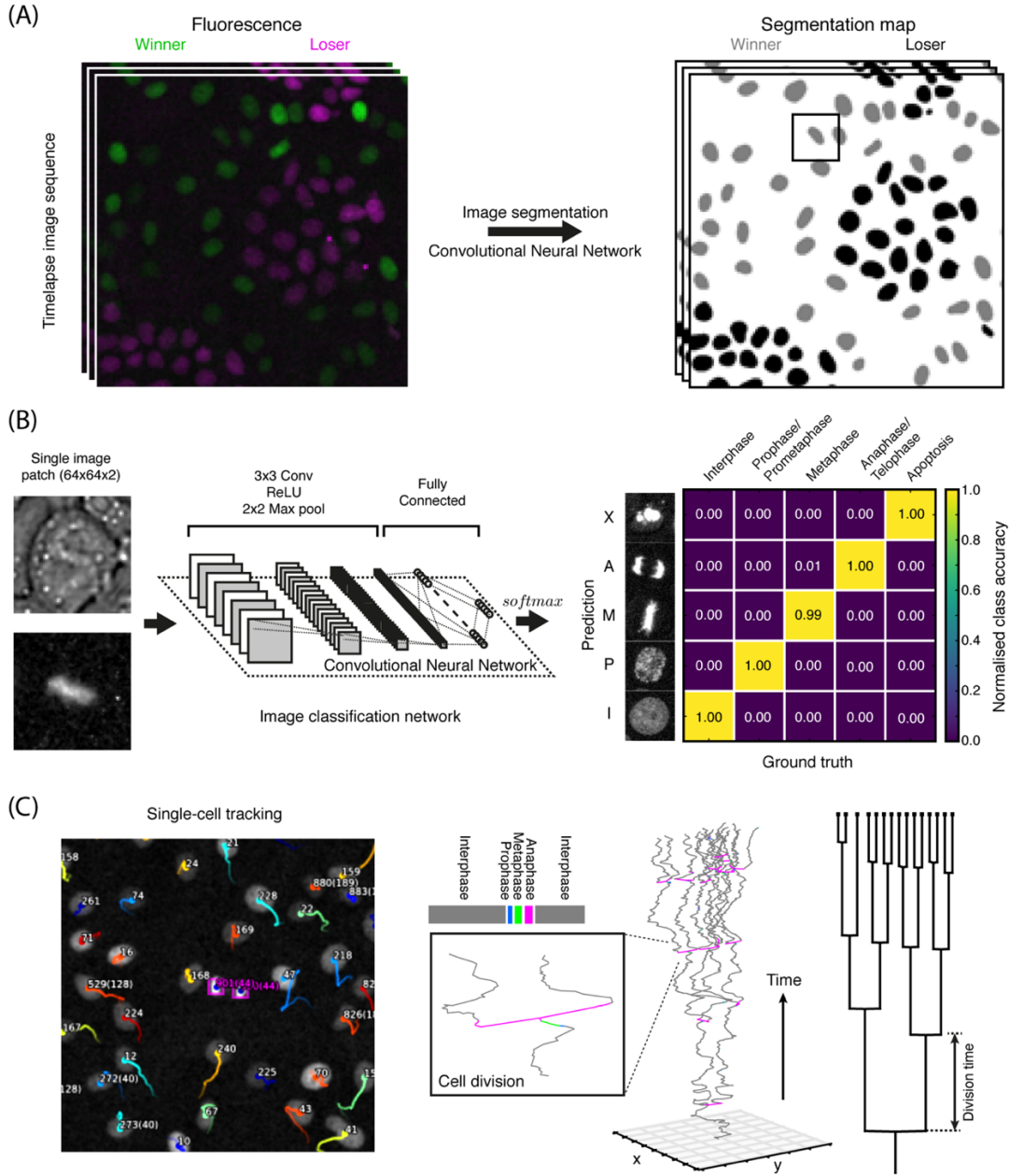

**Figure 2: Experimental workflow for high-throughput analysis of cell competition at the single cell scale.**
(A) High-throughput time-lapse microscopy of mixed cultures of winner and loser cells yields thousands of images per experiment. Each image shows the fluorescent histone markers demarcating cell type (winner or loser). These images can be robustly segmented using computer vision and convolutional neural networks (CNN), allowing the cell type, centroids and cell count to be determined at each time point. (B) A CNN classifier can be trained to accurately identify the mitotic or apoptotic state using image data alone. (C) Using the segmented and classified objects, automated tracking of the cells over time allows the reconstruction of lineage trees. Each lineage tree defines the spatial position, motion, lineage, instantaneous state and fate of every cell in the field of view.

## 2.2 High throughput imaging of cell competition

As competition takes place over durations of tens of hours, long duration movies acquired by time-lapse live cell microscopy are essential to decipher it (**Fig 2**). Considerations when designing imaging assays are setting the spatial and temporal resolutions, as well as a large enough field of view that enables imaging of larger amount of cells. The spatial resolution must be sufficient to easily distinguish the single cell events that underlie competition such as proliferation, death, and migration. The temporal resolution must be such that no event (death or division) goes undetected. If cells are highly migratory, it is imperative to choose a short enough time interval for cell tracking to allow for meaningful measurements of cell velocity or mean-squared displacements. If they are not, the duration of anaphase or cytokinesis will inform on the suitable interval. Having chosen a temporal resolution, one can decide on how many fields of view to acquire based on frame acquisition time.

Analysis of the imaging data is a critical consideration. Experiments typically yield $10^2$-$10^3$ cells per field of view, over hundreds of time-points and multiple fields of view, resulting in millions of cells worth of single-cell data [20]. Fluorescent markers are often used to distinguish winners from losers [19, 20, 24, 25]. Histone markers with different fluorophores are practical because they can be robustly segmented (**Fig 2A**). Although histone markers allow for identification of proliferation and death, they are late indicators of a commitment that has occurred several hours before. Hence, it may be useful to combine histone markers with additional reporter constructs, such as geminin [26] for entry into the cell cycle or active caspase reporters for cell death [27], to gain more precise information on when commitment occurred.

From the image data, each cell must be robustly identified, localized, labelled and tracked over the duration of its existence. Several software packages have been developed for general single-cell tracking. These include *tTt* [28], *CellProfiler* [29], *CellCognition* [30] and *LEVER* [31] amongst others (reviewed in [28]). However, this remains a challenging problem. For example, a method with a perceived high tracking accuracy of 99% actually means that up to 10 cells at every time point are mis-identified or mis-categorized. This is particularly problematic in the study of rare events such as cell death or division (see **section 2.5** for more discussion). Approaches utilizing advanced computer vision and machine learning to perform automated segmentation and event detection (**Fig 2B**) are improving the accuracy of such analysis methods. Our own software, *Sequitr*[1] [20], has been developed for cell competition, combining state-of-the-art deep learning for robust cell segmentation and classification together with a Bayesian approach to tracking (**Fig 2C**).

**2.3 Population and single-cell metrics to characterize competition**
As many different mutations can drive competition, defining general metrics to characterize competition allows us to categorize competitions based on quantitative signatures and determine the effect of perturbations.

Using segmented and classified image data, population fate metrics, such as the probability of cell death/division as a function of time or the cumulative number of deaths/division, can be computed for each cell type. Further insight into competition at the single cell level necessitates correlation of fate with features of the environment likely to influence competition. For example, in comparative fitness competition, losers in contact with winners are eliminated and increased contact leads to increased probability of apoptosis. Therefore, quantifying how proliferation and death depend on the composition of the local neighborhood will likely be informative. Conversely, in mechanical competition, crowding has been shown to be a key variable [19]. Hence, we need to determine the dependency of death and division on local cell density.

---

[1] https://github.com/quantumjot/CellTracking

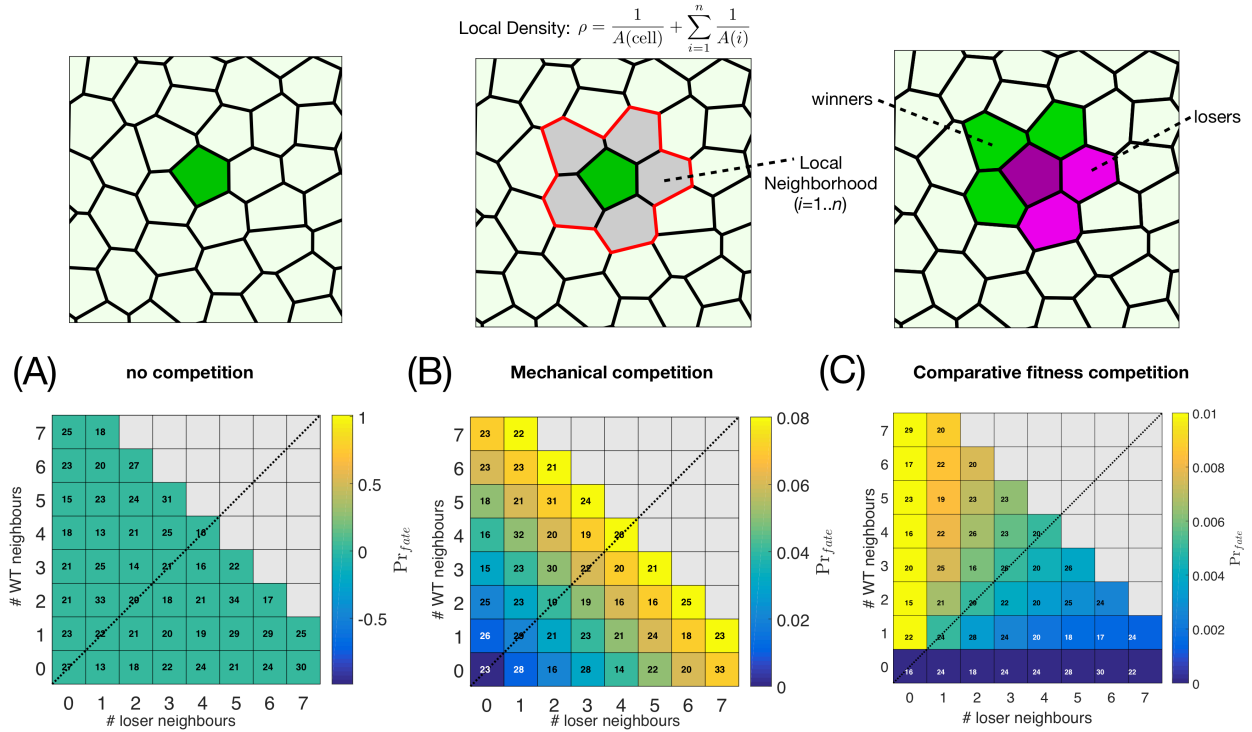

**Figure 3: Single cell metrics to characterize competition.** The mechanisms of competition at the single cell level can be inferred from neighborhood plots (bottom row) which has the number of winner (wildtype – WT) neighbors on the y-axis and the number of loser neighbors on the x-axis, about a chosen cell. The probability of an event (cell fate) is color coded and the number of observations is indicated in each grid position. (A) In a no-competition scenario, the probability of a certain cell fate is independent of the number of the winner/loser cells, resulting in a uniform color map. (B) In pure mechanical competition, the local density of a cell's neighborhood is important, but not the identity of the cell type. In this case, the probability of cell apoptosis will be symmetric about the diagonal, and will increase with increasing number of cells in the neighborhood. (C) In comparative fitness competition, loser cells may possess a higher likelihood of apoptosis when most of their neighbours are winners. As a result, the neighbourhood plot is asymmetric, with higher probabilities of death above the diagonal.

The most obvious definition for cell density is the inverse of the area of the cell of interest. However, this definition can be misleading because some epithelial cells have a similar morphology whether they are alone or surrounded by neighbors. For a more robust definition, the area of the cell and its direct neighbors should be taken into account, for example by defining the density as: $\rho(cell) = \frac{1}{A(cell)} + \sum_{i=1}^{n} \frac{1}{A(i)}$ with *n* the number of neighbors, $A(i)$ the area of neighbor *i*, and $A(cell)$ the area of the cell of interest (**Fig 3B**). Using this definition, a density can be defined for each cell at each time point and events (division, apoptosis, etc) can be binned as a function of density to yield curves relating the probability of the event to the local density. This approach reveals that, in mechanical competition, cell death in loser cells increases sharply with density. Cell death in winners appears to follow a curve similar to the losers but shifted towards higher densities [20]. These observations suggest that some aspects of mechanical competition may be understood through differences in homeostatic density between winners and losers.

In comparative fitness competition, heterotypic contacts govern cell fate and thus a metric that quantifies this is necessary. One simple idea is to categorize cells by their number of winner and loser neighbors. After doing this for each cell and at each time point, we can classify each event and each observation as a function of their winner and loser neighbors to generate a neighborhood plot (**Fig 3A-C**) [20]. This plot has the number of winner neighbors on the *y* axis and the number of loser neighbors on the *x* axis. In each grid position, the probability of the event is color coded and the number of events observed is indicated. This presentation allows to rapidly grasp whether the event of interest depends on the number or identity of neighbors. For example, if there is no competition, the neighborhood plot for apoptosis will have a uniform color: apoptosis does not depend on the number or identity of neighbors (**Fig 3A**). If we are examining mechanical competition, the probability of apoptosis will be higher for cells with more neighbors but their identity is not important. In this case, the neighborhood plot is symmetric about its diagonal (**Fig 3B**). In comparative fitness competition, loser cells may be more likely to die when most of their neighbors are winners. In this case, the neighborhood plot is asymmetric with high probabilities of death above the diagonal and low probabilities of death below the diagonal (**Fig 3C**).

Other metrics may of course be useful in characterizing competition. For example, in comparative fitness competition, the percentage of the perimeter in heterotypic contact and the duration of this contact may be important parameters in determining cell fate. In mechanical competition, stress in the epithelium may play a role. Therefore, computing strain rate from cell velocity maps may be informative.

**2.4 Statistics of rare events**
A major challenge with such metrics becomes evident when we want to compare experiments. The dominating observables in cell competition are rare events, such as cell division or death. How can we determine if a perturbation experiment is significantly different from its control? How can we determine if grid positions are significantly different from one another?

Generally, to compare probability estimators, we determine if their values differ by more than their absolute precision. This absolute precision is often estimated by the standard error: $SE = \sqrt{\frac{p(1-p)}{N}}$ with *p* the estimated probability: *p=n/N* with *n* the number of events and *N* the number of observations. One issue arises from the calculations of statistics of rare events [32]. To illustrate this, consider an event of cell elimination that occurs as an independent Poisson process over the whole duration of a typical competition assay with a constant rate of 1% (0.01 deaths per cell per duration of experiment). Any given cell has a probability 1-$e^{-0.01}$ = 0.00995 of dying over the whole experiment. For a population of 1000 cells observed over the whole experiment, the standard error associated with cell elimination is 0.0031. However, if cells are only observed for 10% of the experiment length, the probability of elimination for one cell becomes 1-$e^{-0.01*0.1}$ = 0.0009995 and the standard error for a population of 1000 cells becomes 0.000992, far less than for the whole experiment. Yet, neither the elimination process nor the cell population have changed. This improvement of precision is an artifact of a rare event. To overcome this issue, one must use measures of relative precision (Coefficient of variation: $cv = \sqrt{\frac{(1-p)}{pn}}$), where the standard errors are calculated relative to the actual probability of the rare event [32]. A $cv < 10\%$ enables robust comparison of grid positions and across experiments. If the $cv$ does not reach such low values, more data needs to be acquired or the existing data needs to be binned into coarser categories.

**2.5 Measuring relevant physical and biochemical variables**
Ultimately, we want to link our observations from competition to underlying differences in physical and biochemical variables. Computer simulations provide a powerful tool to predict cell competition outcomes, test experimental hypotheses and, as input, they require to measure the variables that may play a role in competition for each cell type separately as well as in competitive contexts. For example, for mechanical competition, the strength of intercellular adhesion, the stiffness of cells, and the homeostatic density may all play a role. In comparative fitness or ligand capture competition, the cascade of different biochemical reactions and protein interactions underlying fitness comparison or growth factor uptake would need to be characterized. Methods for measuring each variable are specific to each type of competition and each signaling pathway.

## 3. Computational modelling approaches to reverse-engineer cell competition

### 3.1 Local vs global models of cell competition

The availability of multi-scale quantitative data opens up new possibilities to validate computational models that can provide a bottom-up understanding of the fundamental physical principles governing cell competition.

Models of cell competition can be categorized as global or local depending on whether they are population-based or cell-based, respectively. Global population-based models are usually formulated as ordinary differential rate equations (ODEs) mapping the evolution of the number of winners and losers to continuous variables (**Fig. 4A**). Such models implement phenomenological rules for cell colony growth and decay, without considering local interactions. This simplification enables the use of mathematical analysis to infer the general characteristics of the cell competition phenomena, which can be conveniently tested in population-scale experiments. As attractive as reducing a complex multivariate system to simple differential equations may seem, these models can only predict the emergent bulk properties such as population ratios, their relative growth rates or tissue size. The influence of local topography, mechanics and cell-cell signaling cannot be taken into account, despite experimental and computational evidence showing their importance in metastasis [33] and in cell competition [11, 19, 20, 23]. In particular, ODE models are unsuited to describe morphologies of co-existing colonies observed in comparative fitness competition [11].

ODE based models tend to reduce the multi-variate parameters of cell competition to a single asymmetry parameter for the fitness of competition. Fitness asymmetry is sometimes modelled as the difference in growth rates in an ODE model [6, 34]. Yet, many studies of cell competition have concluded that differential growth is not the only source of competition. Another metric for asymmetry in fitness is the ratio of survival fractions of the two cell populations, formulated using a predator-prey model much like the well-known *Lotka-Volterra* model [35]. In other coupled rate equation models [20], an asymmetry in fitness is incorporated via differential sensitivity of proliferation rate to cell density. Indeed, experiments and modelling showed that differences in apoptosis sensitivity to density alone cannot explain the outcome of competition. One limitation common to all ODE models is that they cannot account for spatial heterogeneities in the population nor single-cell level configurations, topology, and mechanics.

To circumvent these limitations, a common approach is to implement partial differential equation based models as a framework to incorporate active cellular mechanics, motility, as well as couplings between mechanical and chemical fields [36, 37]. Here, the two cell types are modelled as viscoelastic media with pre-defined mechanical and kinetic properties [38-41] where the interaction mechanisms are implemented phenomenologically. A drawback, however, is that such an approach leads to many undetermined physical parameters whose molecular origins remain speculative. Furthermore, the choice of fixed mechanical properties limits the emergent states of self-organization. While continuum models have been used to infer the mechanistic principles of tissue homeostasis [42], they remain ill-suited to describe contact-dependent biochemical competition, as well as topology-based cell-cell interactions that are essential regulators of global population fate.

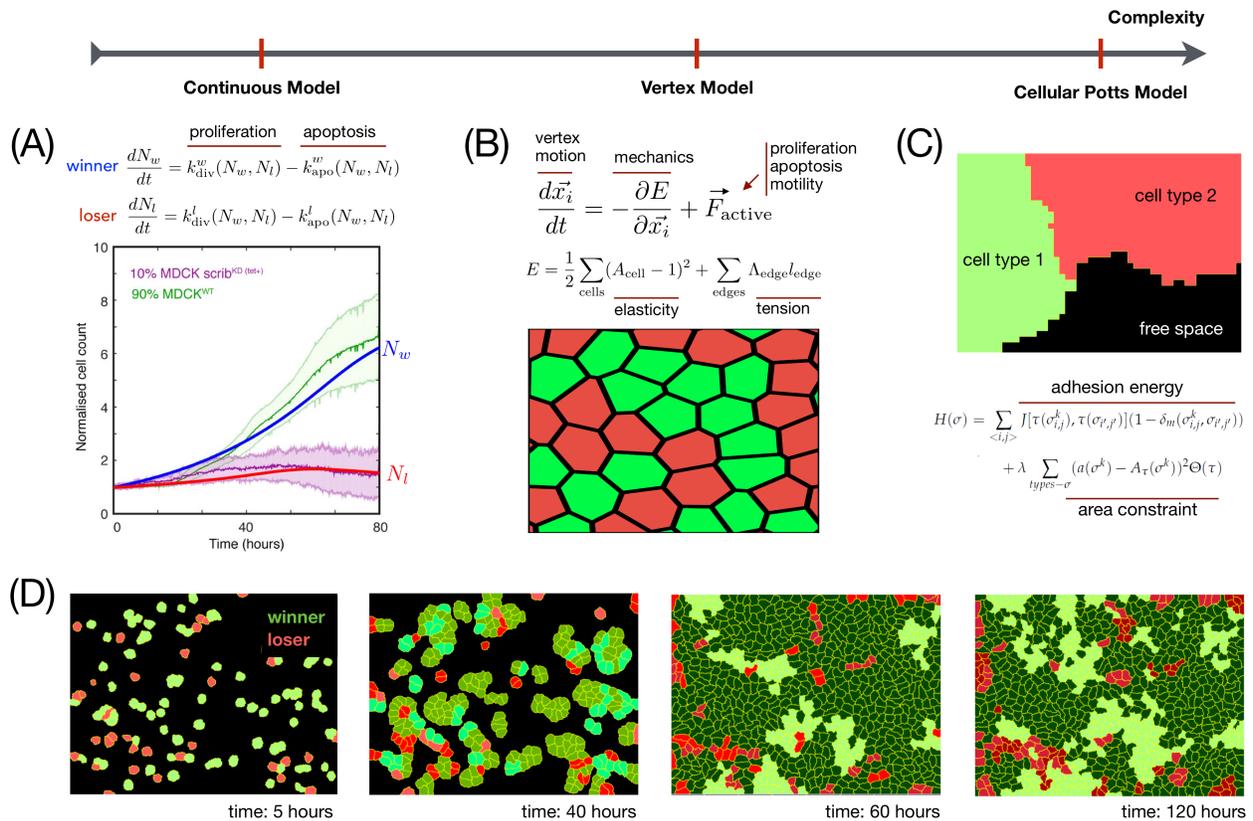

**Figure 4: Computational modeling approaches to reverse-engineer the rules of cell competition.** (A) Continuous model for cell competition based on ordinary differential equations for the rate of proliferation and apoptosis for the winner and loser cells [19]. (B) Vertex modeling of cell competition, where individual cells are represented by polygons in a confluent tissue. Each vertex evolves based on the net force acting on them from mechanical and active processes. (C) Cellular Potts Model for cell competition, where each cell is a sub-lattice of pixels, that can form arbitrary shapes and undergo motile behavior based mechanical forces or stochastic rules. (D) Snapshots of simulated cell competition dynamics using a multi-layered cellular Potts model, where winner cell types are indicated in shades of green and loser cell types are shown in shades of red. Darker shades indicate higher generation number.

### 3.2 Cell-based models

To understand how the behavior of single cells and local cell-cell interactions lead to population-scale dynamics, one needs to explicitly model individual cells as physical entities with interaction rules. These are implemented in the form of computational algorithms using tools from soft matter physics, statistical physics, and dynamical systems.

The vertex model has been extensively used in modelling epithelial tissues due to its conceptual simplicity and computational parsimony. The vertex model treats each cell as a two-dimensional polygon, with vertices and rectilinear edges shared between neighboring cells [43]. Vertices move in response to forces that arise from cell growth, interfacial tension, and intracellular contractile stress (**Fig. 4B**) [44], resulting in changes in cell shape. The choice of polygonal shapes and simple mechanical rules reduce the computational complexity of simulations, while the burden lies in implementing the rules for cell rearrangements and active cellular processes. Vertex models have been widely used in epithelial monolayer modelling including cell sorting [45], germband extension [46], ventral furrow formation [47], cell division and tissue elongation [48], wound healing [49, 50], as well as tumor growth [51]. Vertex models have been used in studies of mechanical cell competition [34], as well as to infer the role of local epithelial topology on mutant clone expansion [23]. Vertex models, however, have some key disadvantages, including the difficulty to model arbitrary cell shapes, isolated cells, or cell clusters. Therefore, this framework is unsuited to capture the competition for free space which is an important aspect of competition prior to confluency.

A common alternative approach is to model cell competition using cellular automata [52], in which dynamic cell behavior is not governed by mechanical force balance but through probabilistic rules for cell apoptosis, mitosis, and mutations. Using this model, Basanta et al have suggested that micro environmental factors such as the local concentration of oxygen or nutrients, and cell overcrowding may determine the expansion rate of the tumor colony. The results also show that tumor cells evolve to adapt their phenotypes to the microenvironment, such that environmental stresses determine the dominance of particular phenotypical traits. A similar approach was used in [33] to model tumor evolution, which showed that short-range cell dispersal and turnover can account for rapid intermixing of cells inside the tumor. How a small selective advantage in a single cell within a large tumor allows colonization of the precursor mass in a clinically relevant time frame is a relevant mechanism for the rapid onset of resistance to chemotherapy. Cell automata models, however, are limited by the fact that they cannot account for the physical interactions between cells known to be important in cell competition and tumorigenesis. These limitations could be overcome using particle-based models [53, 54] that allow for mechanical interactions as well as diverse multicellular configurations.

### 3.3 Cellular Potts Model for cell competition: a multi-layer model.

A comprehensive bottom-up insight into the mechanism of cell competition could be obtained using a combination of rule-based probabilistic models and mechanistic models to implement both signaling and realistic interactions via physical forces. This would result in a multi-layered model where, in its most fundamental layer, individual cells would follow robust mechanical principles governing their physical interactions. The physical layer is then coupled to a layer describing active cellular behavior as rules or (in game theoretical terms) strategies that cells follow to determine their fate. Such multi-layered framework for collective cell dynamics could be implemented following the commonly used Cellular Potts Model (CPM) [55].

The traditional CPM forms the physical layer of the model. At each time point, the CPM ensures that the cell shapes are such that free energy is minimized. The CPM is a mechanical lattice based model (**Fig. 4C**), which represents each cell as a subset of lattice sites sharing the same cell identity. These lattice sites, referred to as cell-sites, constitute the building blocks of the cells in the computational model, and are only assigned the properties of the cell they belong to. For example, each cell type has some value of adhesion energy with other cells/objects as well as a preferred area and a compressibility modulus. The model evolves following a Metropolis algorithm, where a lattice site is chosen at random and assigned a new cell identity. The move is accepted (or rejected) if it lowers (or increases) the value of the Hamiltonian (free energy) function, which is a sum of adhesion energies and volume (or area) constraints. The Hamiltonian can be appropriately modified to implement active cell behaviors such as motility [56].

The CPM enables the addition of an exterior layer comprising the cellular automata rules. These rules are algorithmically executed for each cell separately at each time point to model active cellular behaviors such as growth, mitosis, apoptosis, etc. Such approaches have been applied to the growth of tumor cell clusters by implementing the effects of cell compressibility, active motility and contact inhibition [57]. In the context of cell competition, we have implemented a multi-layered CPM combining cell mechanics with probabilistic rules for cell growth and apoptosis for the two cell types, benchmarked from experimental data. Using this framework, we are beginning to simulate the outcomes of cell competition (**Fig. 4D**) with precise control over physical, biochemical and decision-making parameters.

In summary, modeling approaches that combine local cellular mechanics with active decision-making rules would allow us to test the realism, redundancies and efficiencies of different single cell level strategies governing tissue-scale competition dynamics, and therefore to understand the single cell rules governing competition. With precise control over model parameters, multi-layered simulations would enable us to test the relative roles of each of the physical, biochemical and decision-making parameters on the outcome of cell competition. These outcomes could then be compared to high-throughput experimental data using metrics both at the single-cell level and at the population scale (**Section 2.3**). Such a predictive modeling framework would not only account for bulk physical properties but also predict local properties such as topology, organization, and the fate of single cells.

## 4. Future directions – artificial intelligence and numerical modelling

The bridging of scales from population level and bulk properties to single cell level and mechanism is crucial to understanding the underlying biology and physics of cell competition. Machine learning is now being routinely used to accurately identify and label objects in many forms of image data. We have used such methods to identify and label cells in time-lapse microscopy data of competition [20]. However, the more fascinating question is whether advances in Artificial intelligence (AI) can be leveraged to learn a basic set of deterministic rules of cell competition, as they have already done with chess and go [58] and clinical diagnostics[2]. Could one predict the outcome of cell competition if we knew the rules? For example, which physical, topological and biochemical parameters are important in determining the fate of a single cell? Under what circumstances does fate commitment occur? AI approaches could yield significant insight into these questions by taking advantage of the wealth of image data available to learn general models of behavior, and to uncover parameters which may not be immediately obvious to us. Combined with numerical models that enable us to test the foundations of our understanding by simulating outcomes and testing parameters, these quantitative approaches are likely to yield significant new insights into the mechanisms of cell competition.

---

[2] https://www.ucl.ac.uk/news/2018/aug/artificial-intelligence-equal-experts-detecting-eye-diseases

**Acknowledgements**

We thank Michael Staddon (UCL) for providing help with Figures 1,3-4. This work was supported two Engineering and Physical Sciences Research Council (EPSRC) PhD studentships to DG and AB. SB acknowledges support from UCL Strategic Fellowship, Royal Society and Tata Grant URF\R1\180187. AL and GC wish to acknowledge the support of BBSRC grant BB/A009329/1.